\documentclass[prl,showpacs,preprintnumbers,superscriptaddress,floatfix,twocolumn,amsmath,amssymb]{revtex4}
\usepackage{epsfig}
\usepackage{graphicx}
\usepackage{dcolumn}
\usepackage{bm}
\usepackage[active]{srcltx}
\newcommand{\be}{\begin{equation}}

\newcommand{\ee}{\end{equation}}
\newcommand{\bea}{\begin{eqnarray}}
\newcommand{\eea}{\end{eqnarray}}

\begin{document}

\preprint{}
\title{The Ising model   on a brain network maximizes   Information Transfer
at criticality
  }
\date{\today}
\author{D. Marinazzo}\affiliation{Faculty of Psychology and
Educational Sciences, Department of Data Analysis, Ghent University,
Henri Dunantlaan 1, B-9000 Ghent, Belgium\\}\author{M.
Pellicoro}\affiliation{Dipartimento di Fisica, Universit\'a degli
Studi di Bari and INFN, via Orabona 4, 70126 Bari,
Italy\\}\author{Guo-Rong Wu}
\affiliation{Key Laboratory for NeuroInformation of Ministry of
Education, School of Life Science and Technology, University of
Electronic Science and Technology of China, Chengdu 610054,
China\\}\author{L. Angelini}\affiliation{Dipartimento di Fisica,
Universit\'a degli Studi di Bari and INFN, via Orabona 4, 70126
Bari, Italy\\}\author{J.M. Cortes}\affiliation{Ikerbasque, The
Basque Foundation for Science, E-48011, Bilbao, Spain.}
\affiliation{Biocruces Health Research Institute. Hospital
Universitario de Cruces. E-48903, Barakaldo, Spain.\\}
\author{S. Stramaglia}\affiliation{Dipartimento di Fisica,
Universit\'a degli Studi di Bari and INFN, via Orabona 4, 70126
Bari, Italy\\}

\date{\today}

\begin{abstract}
We implement the Ising model on a structural  connectivity matrix
describing the brain at a coarse scale. Tuning the model temperature
to its critical value, i.e. at  the susceptibility peak, we find a
maximal amount of total information transfer between the spin
variables. At this point the amount of information that can be
redistributed by some nodes reaches a limit and the net dynamics
exhibits signature of the law of diminishing marginal returns, a
fundamental principle connected to saturated levels of production.
Our results extend the recent analysis of dynamical oscillators
models on the connectome structure, taking into account lagged and
directional influences, focusing  only on the nodes that are more
prone to became bottlenecks of information. The ratio between the
outgoing and the incoming information at each node is related to the
number of incoming links.

\end{abstract}

\maketitle


Methods based on the theory of complex networks are becoming more
and more popular in neuroscience \cite{sporns}. Moreover, the
inference of the underlying network structure of complex systems
\cite{barabasi} from time series data is an important problem that
received great attention in the last years, in particular for
studies of brain connectivity
\cite{roudi,bressler,noineuro,friston}. This problem can be handled
by estimating, from data, the flow of information between variables,
as measured by the transfer entropy \cite{schreiber,leh}, a
model-free measure designed as the Kullback-Leibler distance of
transition probabilities. Recently \cite{seth} it has been shown
that transfer entropy is strongly related to Granger causality
\cite{hla}, a powerful and diffuse model-based approach to reveal
drive-response relationships in dynamical systems which is based on
prediction: if the prediction error of the first time series is
reduced by including measurements from the second one in the linear
regression model, then the second time series is said to have a
causal influence on the first one.

In a recent paper \cite{plosone} it has been shown that the pattern
of information flow among the components of a complex system is the
result of the interplay between the topology of the underlying
network and the capacity of nodes to handle the incoming
information, and  that, under suitable conditions, this pattern can reveal the emergency of
the law of diminishing marginal returns \cite{samuelson}, a
fundamental principle of economics which states that when the amount
of a variable resource is increased, while other resources are kept
fixed, the resulting change in the output will eventually diminish.
The origin of such behavior resides in the structural constraint
related to the fact that each node of the network may handle a
limited amount of information. In \cite{plosone} the information
flow pattern of several dynamical models on hierarchical networks
has been considered and found to be characterized by an exponential
distribution of the incoming information and a fat-tailed
distribution of the outgoing information, a clear signature of the
law of diminishing marginal returns. This pattern was thus found in artificial hierarchical networks, and in
electroencephalography signals recorded on the scalp.

Motivated by the evidence that brain function resides in its ability
to process and store information, and by the fact that brain
dynamics is associated to criticality \cite{fraiman,chialvo,taglia},
in this study we investigate a dynamical model implemented on a
structure derived from the actual structural connections in the
human brain. We consider an anatomical connectivity matrix
describing the brain at a coarse scale (66 nodes), obtained via
diffusion spectrum imaging (DSI) and white matter tractography
\cite{hagmann}, provided by one of the authors of the original
paper. We implement on it an Ising model with Glauber dynamics
\cite{glauber}, estimating numerically  the information transfer
between spins. Varying the temperature, the susceptibility shows a
peak which is related to a phase transition occurring in the limit
of large networks \cite{doro}, characterized by long range
correlations; although we are dealing with a network of small size,
we will refer to the temperature at the peak of $\chi$ as the {\it
critical} state of the system. We find that at criticality the Ising
model dynamics results in the maximal amount of total information
transfer among variables and that this information transfer is
affected by the law of diminishing marginal returns, as it appears
comparing the distributions of incoming and outgoing information.
The spatial modulation of this phenomenon is analyzed evaluating, at
each node, the  ratio $r$ between the outgoing and the incoming
information. It turns out that $r$ is related to the in-strength of
the brain network: nodes with high $r$ are those more prone to
become bottlenecks as the information flow increases.

The couplings of the Ising model are $J_{ij}=\beta A_{ij}$, where
$A$ is the $66\times 66$ anatomical connectivity matrix, which
corresponds to an undirected weighted network with average degree
17.4. Whilst in \cite{plosone} we studied the diluted Ising model
on an artificial network, here we analyze the Ising model on the
structural architecture of the brain, characterized by two main
modules corresponding to the two hemispheres. Since it has been
shown that for Ising models Granger causality provides a good
approximation to the transfer entropy while being computationally
much more efficient \cite{isingpellicoro}, here we estimate
information flows in terms of the Granger causality; in other words,
we adopt the Gaussian approximation to the Ising model. The order of
the regressive model (maximum lag) is  fixed to $m=1$. Samples of
$10^4$ iterations are used to estimate the Granger causality; we
verified that these samples are long enough to provide robust
results.
 Moreover, we have also estimated the total bivariate transfer entropy (summing over
all pairs of spins connected by a non-vanishing interaction) and
found very similar results to those obtained by Granger causality. We
remark that direct evaluation of the multivariate transfer
entropy is feasible only for very small systems; a promising
approach, which might render larger systems tractable, is described
in \cite{lik} where transfer entropy is expressed as a likelihood
ratio.
\begin{figure}[ht!]
\begin{center}
\epsfig{file=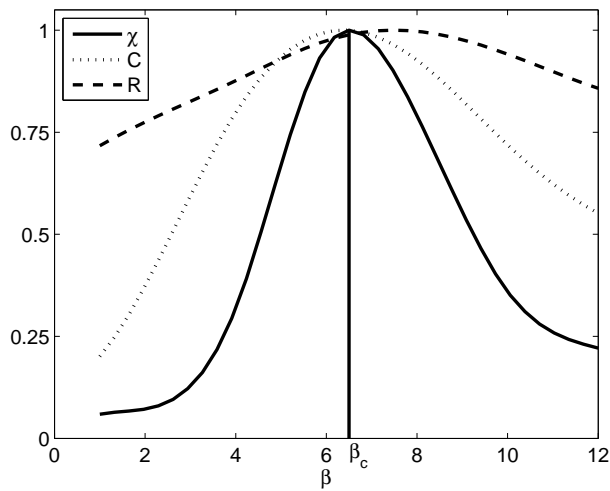,height=7.cm}
\end{center}
\caption{{\small For the Ising model implemented on the brain network,
the following quantities are depicted versus the temperature
$\beta$: R, the ratio between the standard deviations of outgoing
and incoming information transfers ; the total transferred information C, i.e. the sum
of all information transfers in the network; the susceptibility $\chi$.
All quantities have been normalized in the interval
[0,1].\label{66_2}}}
\end{figure}


\begin{figure}[ht!]
\begin{center}
\epsfig{file=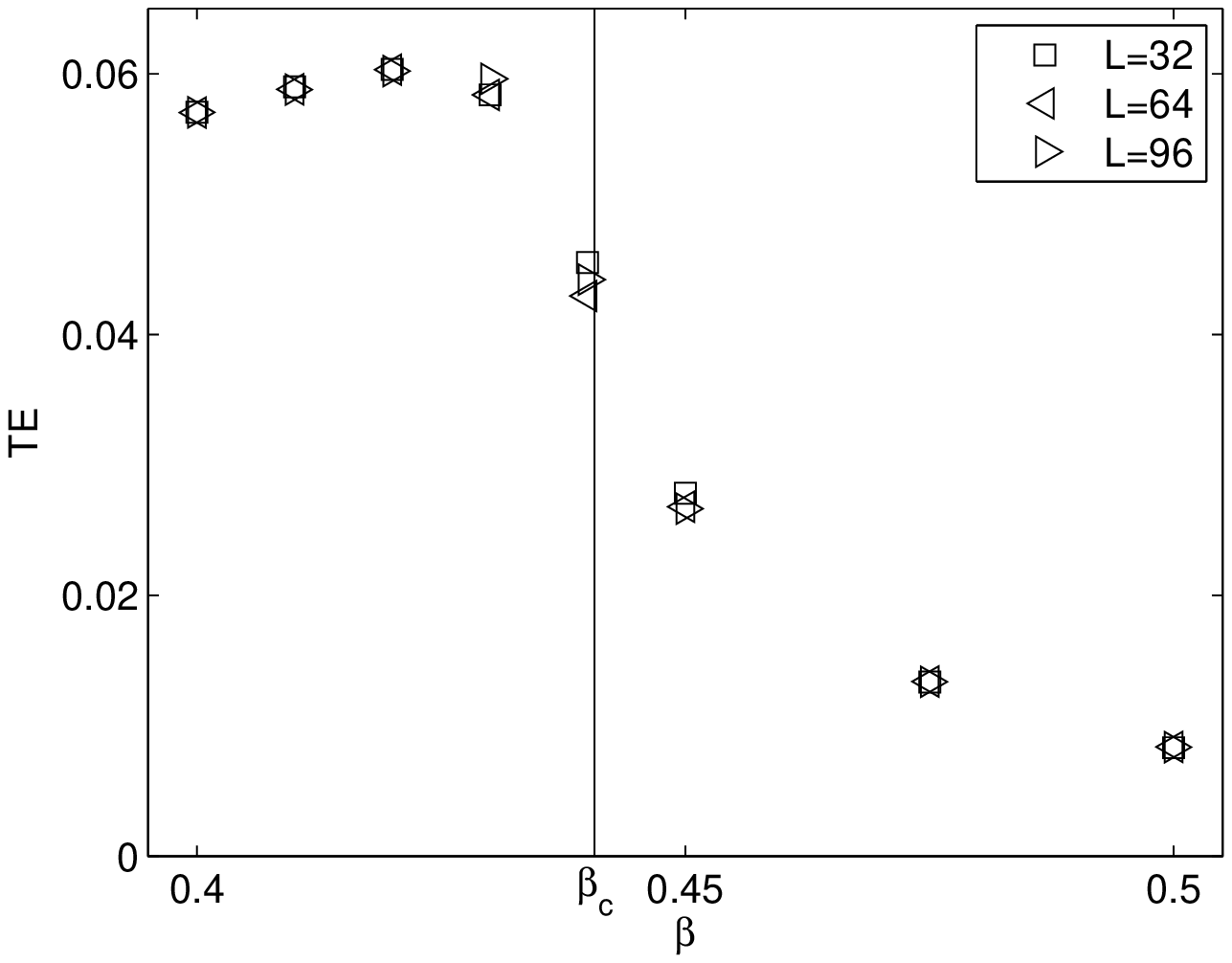,height=7.cm}
\end{center}
\caption{{\small The sum of bivariate transfer entropies for all
pairs of spins, connected by a non-vanishing interaction, is
depicted versus the coupling $\beta$ for the 2D Ising model on a
square lattice of size $L^2$, with $L=32,64,$ and $96$, with
periodic boundary conditions. Transfer entropies have been evaluated
averaging over 20 runs of 10000 iterations. The vertical line
corresponds to the critical point. Since the maximum of the heat
capacity matches quite well the critical point, this suggests that
the system of linear size 96 is a reasonably good approximation to
the thermodynamic limit, and that the transfer entropy has its
maximum in the paramagnetic phase. \label{2d}}}
\end{figure}

\begin{figure}[ht!]
\begin{center}
\epsfig{file=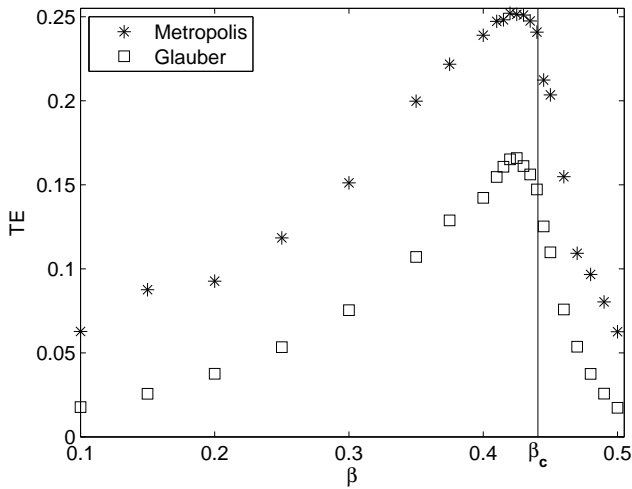,height=7.cm}
\end{center}
\caption{{\small The sum of bivariate transfer entropies for all
pairs of spins, connected by a non-vanishing interaction, is
depicted versus the coupling $\beta$ for the 2D Ising model on a
square lattice of size $L^2$, with $L=16$, with periodic boundary
conditions. Transfer entropies have been evaluated averaging over 20
runs of 10000 iterations. The vertical line corresponds to the
critical point. Stars refer to Metropolis updating scheme, whilst
empty squares refer to Glauber dynamics. \label{2d1}}}
\end{figure}

In figure (\ref{66_2}) we depict three quantities as a function of
the inverse temperature $\beta$: (1) the ratio between the standard
deviation of the distributions of the outgoing information and the
incoming information \cite{plosone},
$$ R={\sigma   (c_{out}) \over \sigma   (c_{in})},$$
where $c_{out}$ ($c_{in}$) is obtained by summing over columns
(rows) the matrix $c = \{ c_{ij}\}$ of the information flows $i \to
j$ as estimated by Granger causality. As explained in
\cite{plosone}, $R$ is an indicator of the law of diminishing
marginal returns, and it is a quantity calculated in this case at a
global level, pooling all the nodes together (complementary to the
local measure $r$); (2) the sum $C$ of all the information transfers
between spins, quantifying the circulation of information in the
network, obtained summing matrix $c$ over rows and columns; (3) the
susceptibility $\chi$, whose peak corresponds to criticality
(pseudo-transition). We find that the around this critical state of
the Ising model the total amount of information transfer
and R assume large values; while the total transferred information is maximum at the critical point, $R$ is maximized for a lower temperature. At criticality some units
are close to be receiving the maximal amount of input
information, but the explanation of why these two related quantities are maximal for different temperatures will need further investigation.


It is worth mentioning that the transfer entropy may be seen as a
dynamical counterpart of the mutual information, the static measure
of statistical dependencies among components of the whole system. It
is well known that for a large class of dynamical systems that the
mutual information peaks at the order-disorder phase transition
\cite{gu,lau}.
We find that for the 2D Ising model the transfer entropy has a peak
in the paramagnetic phase, see figure (\ref{2d}). Moreover we stress
that the amount of information flow depends on the updating scheme,
but the maximum is attained in correspondence of the same coupling,
see e.g. figure (\ref{2d1}) where Metropolis and Glauber dynamics
are compared for the 2D Ising model.

\begin{figure}[ht!]
\begin{center}
\epsfig{file=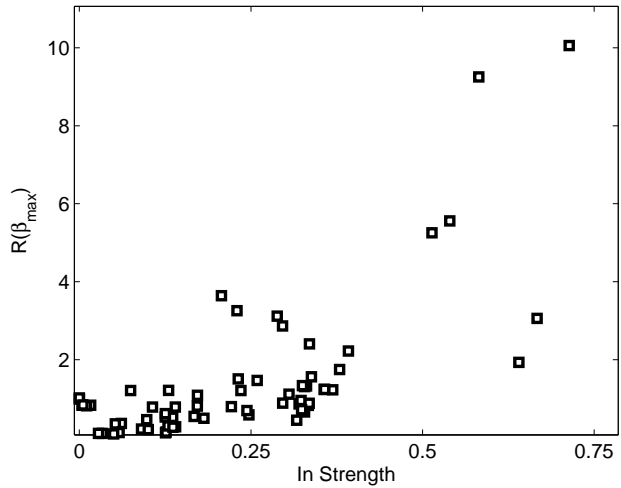,height=7.cm}
\end{center}
\caption{{\small The ratio between the outgoing and the incoming
information at each node is depicted versus the in-strength of
nodes. Pearson linear correlation is  0.69. Varying $\beta$, the
pattern remains the same although the numerical values of r slightly
change. \label{66_4}}}
\end{figure}

The modulation of the law of diminishing marginal returns can be
analyzed evaluating, at each node, the ratio between the outgoing
and the incoming information.
\begin{figure}[ht!]
\begin{center}
\epsfig{file=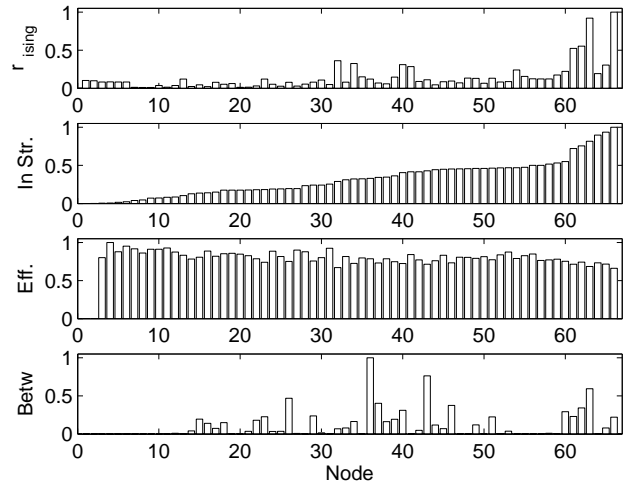,height=7.cm}
\end{center}
\caption{{\small The pattern of $r$ for the Ising model is shown and
compared with the in-strength, i.e. the indegree in the weighted
connectivity matrix $A_{ij}$, the efficiency and the betweeness of
nodes \cite{barabasi}. The values of r have been normalized to the
interval  [0,1]. Nodes are ordered according to increasing
in-strength. Note that r is not fully explained by the in-strength,
there are node with intermediate in-strength but with high r: the
ratio r thus measures a property of nodes which is connected but not
equivalent to the in-strength. Similar patterns are obtained at
critical $\beta$. \label{66_3}}}
\end{figure}
Figure (\ref{66_3}) refers to the value of $\beta$ leading to the
maximum of $R$, and describes, for each node, the ratio
$$r={\langle c_{out}\rangle \over \langle c_{in}\rangle} $$ compared with topological properties of the
graph, such as the strength, the node-efficiency and the
node-betweeness; nodes have been ordered according to growing values
of the in-strength, i.e. the number of incoming links. In figure
(\ref{66_4}) we plot $r$ versus the in-strength, for the Ising
model. We observe that $r$ is to some extent correlated by the in-strength
and it thus reflects an intrinsic property of nodes, the propensity
to become bottlenecks of information. Fig. (\ref{66_3}) shows that
other network properties related with "hubness" such as efficiency
(connected with the shortest path length between neighbors of a
given node, when that node is not present \cite{barabasi}) and
betweeness (accounting for the number of shortest paths passing
through a given node) are not correlated with the the r value in
each node.

\begin{figure}[ht!]
\begin{center}
\epsfig{file=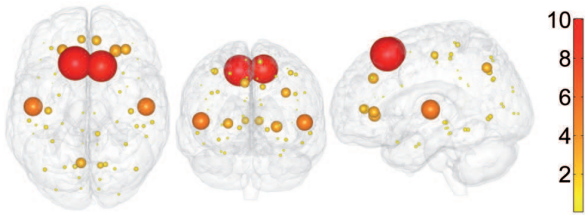,width=8.cm}
\end{center}
\caption{{\small The value of r for each of the 66 regions for the
Ising model. The size of the spheres is proportional to r, thus
showing the most prone regions to became bottlenecks of
information.\label{66_5}}}
\end{figure}

The regions which are recognized as potential bottlenecks by the
present analysis are symmetrical in the two hemispheres, and are
Superior Frontal Cortex, Precuneus, Superior Temporal Cortex, Medial
and Lateral Orbitofrontal Cortex, see fig.(\ref{66_5}). Some of
these regions are considered as hubs both for the structural and for
the functional connectome. It is worth to recall anyway that being a
hub (in particular for incoming connections) does not necessary
imply that a node is a bottleneck of information flow.

%

Recent works \cite{fraiman,cabral,deco,deco1,taglia} have simulated
the spontaneous brain activity implementing models of dynamical
oscillators with different levels of complexity and biological
foundation on the connectome structure, retrieving in some cases
correlation-based networks similar to those observed from the
analysis of neuroimaging data (mainly fMRI at rest). The present
work extends the analysis to dynamical networks who take into
account lagged and directional influences. We have shown that the
critical state of the Ising model on a brain network, unlike the
Ising model on a regular 2D grid, is characterized by the maximal
amount of information transfer among units, and that brain effective
connectivity networks may also be considered in the light of the law
of diminishing marginal returns: some units more prominently express
this disparity between incoming and outgoing information and are
thus liable to become bottlenecks of information.

We have also implemented the Ising model on a symmetric network
built by preferential attachment \cite{barabasi}, and with the same
weight on links. As $\beta$ is varied, we find that the peak of the
susceptibility does not correspond to the peak of the information
flow.  This confirms what we have already found in relation with the
2D Ising model, i.e. the characterization of the critical state as
the maximizer of the total information transfer is not a generic
property; on the other hand it suggests that the distribution of
weights of the links, rather than the heterogeneity, is crucial for
the connectome in order to show such characterization.

Apart from the insights on how structure and dynamics interact to
generate brain function, the approach here described could have more
general implications revealing nodes of a network which are
particularly representative \cite{Liu2013} or influential for the
others \cite{Liu2011}.


\end{document}